\documentstyle[epsf]{article}
\begin{document}

\begin{center}
{\large \bf Physical Results by means of CompHEP}

\end{center}

\begin{center}
{\bf P.A Baikov, E.E. Boos, M.N. Dubinin, V.F. Edneral, V.A. Ilyin, \\
D.V. Kovalenko, A.P. Kryukov, A.E. Pukhov, V.I. Savrin, A.V. Semenov,
S.A. Shichanin }\\
Institute of Nuclear Physics, Moscow State University \\ 
119899 Moscow, Russia 
\end{center}
\begin{abstract}

The CompHEP package was developed for calculations of decay and high
energy collision processes with, correspondingly, up to 5 and 4
final particles in the lowest
order (tree) approximation. The main idea put into CompHEP was to make
available passing from the Lagrangian to  final distributions 
efficiently with high level automation what is extremely needed in 
collider physics.

The present talk describes a general structure of the CompHEP facilities
and reports some physical results obtained with its help. The main purpose
of the talk is to attract the attention of high energy physicists to
this user-friendly package which is completely aimed at making easier
 their routine and tedious calculations in the TeV region. 
\end{abstract}

\begin{center}
{\bf  1. Introduction}
\end{center}

 CompHEP project was started in 1989 by group of physicists and programmers  
 from the Institute of Nuclear Physics, Moscow State University. 
   The first versions of the CompHEP package were written in Turbo Pascal for 
IBM compatible PC. In 1992 this package was rewritten in the C programming
language and now the installation on UNIX workstations is available.
At present time there are some  versions for  different platforms: HP Apollo 
9000, IBM RS 6000, DECstation 3000, Sparc station, Silicon Graphics.
          
      CompHEP is a menu-driven system with the mixed text/graphical output 
  of information and the context HELP facility. The notations used in CompHEP
are very similar to those used in particle physics. It contains
several buit-in theoretical models of particle interactions including the
Standard Model in the unitary and 'tHooft-Feynman gauges. A
creation of a new   particle 
 interaction model by the user is available. The user can change
 interaction vertices and model parameters.  In the present version   
 polarizations are not taken into account. Averaging over initial and summing 
 over final polarizations are performed automatically.

   The general structure of the  CompHEP package is represented  in Fig. 1.
It consists of the {\it symbolic} part and the {\it numerical}
one. The main facilities of the symbolical part allow the user to        
\begin{itemize}
    \item
 select process  by specifying  {\it in-}  and {\it out-} 
particles for decays of $1 \to 2, \ldots ,1 \to 5$ types and
collisions of  $2 \to 2, \ldots, 2 \to 4$ types;
\item
  generate and display tree-like Feynman diagrams in the lowest order;
\item
 eliminate some number of diagrams from the further consideration; 
\item
 generate and display squared Feynman diagrams (corresponding
to squared $S$-matrix elements);
\item
derive analytical expressions corresponding to squared 
  diagrams with the help of the fast built-in symbolic calculator;
\item
  carry out numerical calculations  for $1 \to 2,\; 1\to 3$ and $ 2 \to 2$
processes and show plots of various distributions on the screen;
\item 
generate LATEX files for graphical outputs;
 \item         
   save analytical results  in the REDUCE and MATHEMATICA codes for further 
    symbolical manipulations;
\item
  generate the optimized FORTRAN codes for the squared matrix
 elements in order to make  further numerical calculations.
\end{itemize}

     The numerical part of the CompHEP package is written in  FORTRAN. It 
uses 
     the CompHEP FORTRAN output and the  BASES\&SPRING package \cite{r2} for 
Monte-Carlo integration and event generation. By means of the
CompHEP numerical  part the user is able to 
\begin{itemize}
 \item      
      choose phase space kinematical variables;
    \item
     introduce kinematical cuts on  momentum transfers  and 
    squared masses for any groups of outgoing particles;
    \item
     make  regularization to  remove  sharp peaks from the  
integrand;
     \item
     change the BASES parameters of the  Monte-Carlo integration;
\item
     change numerical values of  model parameters;
    \item
     calculate distributions, cross sections or 
    particle widths by the Monte-Carlo method;
\item
     carry out integration taking  account of structure functions 
        of incoming particles;
 \item       
     generate events and  obtain histograms simulating a signal
       in the real experiment.

\end{itemize}

\begin{center}
{\bf  2. Menu system of the CompHEP  symbolic part }
\end{center}

The user  can select  menu positions displayed on the screen
     with the help of the {\tt Arrow} keys. 
     The input of the selected position is performed by pressing the 
     {\tt Enter} key.
     One can press the {\tt F1} key for {\sf Help}, i. e. in order to get 
     information about the selected menu position. 
     To  return to the previous level menu  one should  click the {\tt Esc} or 
{\tt Backspace} keys.

 The menu titles of the CompHEP symbolical part are shown in Fig.2.

{\it Menu 1} (models)

This menu gives the user a possibility to select a model  of 
elementary particle interaction. 
 {\sf  Fermi model}   includes  QED and the four-fermion weak interaction. 
         The interaction of fermion currents is implemented through 
          the auxiliary intermediate bosons with constant
          propagators.               
     
 {\sf N E W \quad M O D E L} is an option for creating a new
       physical model. The user  will be asked
about a new model name and a template for the model. To choose the
template the list  of the existing models appears. The {\sf Edit model} 
       option of {\it Menu 2} can be used to insert changes. 

{\it Menu 2}

 {\sf Enter process} is an option for entering the process from the
keyboard with the CompHEP notations displayed on the top of the
screen as a table: 
           
\begin{center}
\begin{tabular}{|l|l|l|}	  
\hline
A(A) -photon &  G(G) -gluon & e1(E1) -electron \\
n1(N1) -e-neutrino  & e2(E2) -muon & n2(N2) -mu-neutrino \\
e3(E3) -tau   &  n3(N3) -tau-neutrino & u(U) u-quark \\
d(D) -d-quark &  c(C) c-quark & s(S) s-quark \\
t(T) t-quark  &  b(B) b-quark & H(H) -higgs \\
W+(W-) W-boson & Z(Z) Z-boson   &  \\
\hline
 \end{tabular}
 \end{center}

    If the  input is correct CompHEP constructs the 
corresponding Feynman diagrams and the user gets to  {\it Menu 4}.

    The {\sf Edit model } option leads the user to the {\it Menu
3} to insert changes into model tables.

{\it Menu 3} (edition of models)

 A physical model in CompHEP consists of four tables:
{\sf Parameters, Constraints, Particles} and {\sf Lagrangian}.  
For example, the latter looks like the following:
{\small
\begin{center}
\begin{tabular}{|l|l|l|l|l|l|}   
\hline
   A1 & A2 & A3 & A4 & Factor         & Lorentz part            \\
\hline
W+ &W- &Z  &   &EE*CW/SW          &m1.m2*(p1-p2).m3+m2.m3*(p2-p3).m1 \\
   &   &   &   &                  &+m3.m1*(p3-p1).m2 \\
N1 &e1 &W+ &   &EE/(2*Sqrt2*SW)   &G(m3)*(1-G5)                           \\                    
E1 &n1 &W- &   &EE/(2*Sqrt2*SW)   &G(m3)*(1-G5)                           \\                  
U  &u  &G  &   &GG                &G(m3)                                  \\                
 ...&...&... &... &...                     &...                    \\   
\hline
\end{tabular}
\end{center}
}

 After inserting the changes in the tables  
 CompHEP  checks the  new version of the model. If the version is correct 
it is saved into the user's directory  {\sf models }.  Otherwise the 
message about an error appears on the screen.
 
{\it Menu 4 }
 
 {\sf View diagrams}  displays  graphical  presentation  of the constructed
Feynman diagrams (see Fig. 3). Here the user has a possibility
to exclude some diagrams from further processing. The LATEX
output of any diagram can be generated automatically.

{\sf Squaring }  generates  diagrams for squared $S$-matrix elements.
                 
{\it Menu 5}

Here the user can {\sf View squared diagrams } and again has a possibility to 
exclude some diagrams from a further consideration. 

    In CompHEP the {\sf Symbolic calculations} of the generated squared
diagrams are performed by means of the built-in symbolic manipulation package.
The user can {\sf Write results } for squared diagrams on the
hard disk in different formats (see {\it Menu 6}), the
corresponding files are placed in the directory {\sf results}. 

  {\sf REDUCE program }  generates  
 source codes for the following calculation of the squared
matrix element by means of the REDUCE package.  
		  
The built-in {\sf  Numerical calculator} fulfills calculations for the 
simplest $1 \to 2$ and $2 \to 2 $ processes. Calculated
numerical values of widths or cross sections for a given process 
are displayed on the screen. It can calculate various
distributions as well (see {\it Menu 7}).

  {\sf Interface} gives a possibility to incorporate the CompHEP session 
with the work of other external packages. The menu of external 
packages appears. The first position of this menu is used to start the 
numerical part of the CompHEP package (see Section 3).  
                   
{\it Menu 7}

 {\sf  View/change data} shows the table with parameter names 
and their numerical values
on the screen.  After inserting the changes the cross section
(width) will be recalculated  automatically.

    The user can {\sf Set angular range}  putting in
 the \verb!min! and \verb!max! values of cosine of scattering
angle in the center of mass  reference frame.
It is also available to {\sf Set precision }  of numerical calculations. 

One can calculate the {\sf Angular dependence} 
(differential cross section) of the scattering process
in the center of mass system.  Through the {\it Menu 8} 
the user can urge CompHEP to {\sf Show plot}, {\sf
Save results in a file} and generate the LATEX file of angular 
distribution. The files will be saved in user's directory
{\sf results}.

    The {\sf Parameter dependence } option allows one using the
{\it Menu 9} to get distributions
for  cross section (width) and asymmetry in any range of parameters
relevant to the process under consideration. 
The {\it Menu 10} should be used to get tables and pictures for
any distributions. 

    In the recent version of CompHEP the {\sf Dalitz plot} facility
    has been implemented for decay processes of $1\to 3$ type. It provides 
    a possibility to investigate    event distributions within phase space 
    and to get vertical and horizontal slices of the Dalitz plot.

\begin{center}
{\bf 3. Numerical part of the CompHEP package}
\end{center}

This facility allows the user to prepare the
CompHEP FORTRAN output for a further numerical integration over 
phase space and  to carry out this integration in a user-friendly
manner. It provides also an interface with the 
Monte Carlo integration package BASES and the 
event generator SPRING.
  Thus the user has a possibility to calculate 
decay rates, collision  cross sections and 
 fill  in histograms for various distributions.
However the package is unable to make a
     summation over types of {\it in-} or {\it out-} particles.

     There are available two run modes: the interactive
     and batch ones. In the interactive mode the package is a menu-driven
system.
The structure of the CompHEP numerical part is reproduced in
Fig.4. It consists of the {\it Main Menu} and a set of submenus.
To select the menu position the user should  type its number and press 
the {\tt Enter} key.  To get HELP relevant to any menu the user should enter
\verb!h#! where \# is a position number or \verb!h! for
general HELP.

    The {\it Calculation} position of the {\it Main Menu} starts
 calculation of the collision cross section 
or the decay width by BASES. 
The program  operation is organized as a sequence of  working 
sessions with a  displayed number automatically increasing after 
each sequential  Monte Carlo calculation.
    Other {\it Main Menu} positions call the submenus for
setting the environment of Monte Carlo integration.
Below we give a brief description of the submenu titles.

The {\it IN state} submenu  serves for preparing the initial 
state of collision processes and allows to:
  \begin{itemize}
  \item enter the CMS energy of {\it in}-particles; 
  \item  switch on  structure function options.
  \end{itemize}

  The {\it Model parameters} submenu
     allows one to change any physical model parameters 
relevant to the studied process and save the new values in 
 a file.

The {\it Invariant cuts} submenu 
   is used to introduce kinematical cuts 
   on  any squared
combinations of Lorentz momenta of {\it in-} and {\it out-}particles.
These cuts are written down in a table like the following:

\vspace{.5cm}
\vbox{ 
\begin{tabular}{|c|c|c|c|c|}
\hline
         N & MIN VALUE &  INVARIANT           & MAX VALUE & STATUS \\
           &  [GeV**2] &                      &  [GeV**2] &        \\
        \hline 
         1 &           &  (-p1+p3)**2         & $<$ -1.000 & HARD   \\
         2 &           &  (-p2+p4)**2         & $<$ -1.000 & HARD  \\
        \hline 
\end{tabular}   
}

\vspace{.5cm}

  One can use the {\it Kinematics} submenu  for defining integration
variables. This submenu allows the user to define
kinematics in the most proper way for
the further integration over phase space.  An algorithm of
this parametrisation is
based on the idea that any process can be considered as subsequent
kinematical decays of particle clusters into two other subclusters. 
The  scheme of kinematical variable selection are fixed in a table
which, for example, in the case of the $2\to 3$ process might
take the following form:

\vspace{.5cm}
\vbox{ 
\begin{tabular}{|c|c|c|c|c|}
\hline
Cluster &    In    & Out 1  & Out 2 & Pvect \\
\hline
 1  &   p1+p2  &   p3  &  p4+p5   &  -p1  \\           
 2  &   p4+p5  &  p4   &   p5     &   -p2 \\
\hline
\end{tabular}   
}
\vspace{.5cm}
\noindent In the last column \verb!Pvect! is chosen for fixing 
polar coordinates in such a way that the integration variables
coincide with the variables in which the integrand is singular.

  The {\it MC parameters} submenu allows 
the user to change some BASES  parameters which are engaged  in this
package.   There are  two loops of the BASES calculation.
 The  first one consists of iterations
  with  grid adaptation  from iteration to iteration.   
The second loop includes iterations with the fixed grid to
accumulate necessary statistics. So the user should define:

\begin{itemize}
\item number of Monte-Carlo sample points for one iteration; 
\item maximal number of iterations with  grid 
adaptation (1st loop);
\item  limit of the calculation accuracy in \% (1st loop);
\item maximal number of iterations with the fixed grid (2nd loop); 
\item limit of the calculation accuracy in \% (2nd loop).
\end{itemize}

Also it is possible here to  switch on the event generator SPRING.
After completing  the BASES calculation the {\it Event generator} 
submenu (the
interface with the SPRING package)  appears
if the event generator has been  switched on:

\begin{center}
\begin{tabular}{|ll|}
\hline
 \multicolumn{2}{|c|}{Event generator menu } \\
  \hline 
    1: Start generator  &  2: Number of events = $10000$  \\ 
    3: View current `hst` file  &                  \\  
\hline                                    
\end{tabular}
\end{center}

The file {\sf hst.\#} contains the report of the SPRING run with histograms 
initialized  by the user.

 The {\it Regularization} submenu is used to 
transform integration variables
for representing the  integrand as a smooth function
in the cases when the squared matrix element has 
singularities and 
 the  Monte Carlo integrators are not efficient enough.
  For a reliable evaluation of such singular integrals the package has  
  special
options which can be activated in the submenu.  Certainly this option is
available only if there is the correspondence between singularities and a
set of the integration variables. So to make the regularization it could be
necessary to change kinematics.
The invariants over which the regularisation is made are written down
in a table. For example, in the case of $t$-channel singularity
it might look like this: 
 
\vspace{.5cm}
\vbox{ 
\begin{tabular}{|c|c|c|c|c|}
\hline
 N &  INVARIANT  &  MASS [Gev]  &  WIDTH [Gev]   & STATUS \\
\hline 
    1 & (-p1+p3)**2         &  0             &             &  ON    \\
    2 & (-p2+p4)**2         &  0             &    &  ON    \\
\hline
\end{tabular}   
}
\vspace{.5cm}
 The {\it Task formation} submenu provides the
following options: 

\begin{itemize}
\item  to collect results of the calculation in table(s) with any physical
   parameters as a table agrument;
\item  to prepare a task for batch mode calculation;
\item  to set default session parameters.
\end{itemize}

    The {\it User's menu} serves for implementing some
user's routines allowing, for example, to:

\begin{itemize}
\item  introduce cuts for any functions of kinematical variables;

\item   convolute squared matrix elements with 
     any structure functions, for example, from the CERN PDF library.

\end{itemize}

 The {\it View results} submenu allows the user 
 to view any output files containing results of
   cross section (width) calculations, a report on the MC
   integration process, histograms.
As a result of the calculations for each working session the
program creates three  output files (\# denotes a session number):

\begin{itemize}
\item {\sf res.\#} contains a result of the calculation with a list of model
  parameters used.

\item {\sf prt.\#} is a copy of the screen report of 
  calculation with a list of all parameters (technical and physical ones).

\item {\sf hst.\#}  contains filled in histograms.
\end{itemize}

\begin{center}    
    {\bf 4. Brief review of physical results \\
     obtained by means of the CompHEP package}	 
\end{center}

    Ten three-body processes in the $e^+e^-$ collisions for a heavy particle
       production 
such as Higgs boson, $t$-quark, $W$- and $Z$-bosons are
calculated in Ref.\cite{r3}
 by two independent computer codes (generated by CompHEP\cite{r1} and 
GRACE\cite{r2}).
The results are in an excellent 
agreement
within statistical errors of numerical integration (about 0.5\%). This
 cross-check of numerical results demonstrates that the
CompHEP and GRACE systems are quite reliable for a theoretical study of 
processes at future 
$e^+e^-,\; e\gamma$ and 
$\gamma\gamma$ colliders.
Various $2\to 3$ reactions for the Higgs production in association
with a vector boson pair at future $e^+e^-$ colliders are calculated 
also  in the paper
 \cite{rCuypers} using the amplitude technique and the CompHEP  package. 
A very good agreement
of two independent calculations has been found. The paper demonstrates
an important feature of CompHEP application as an additional test
of results obtained by other methods or computer systems.

Cross sections of the Higgs boson associated production in $\gamma e$ 
collisions
are calculated in Ref.\cite{r4} for the $\gamma e\to \nu W H$ and 
$\gamma e\to
e Z H$ processes. Event signatures for the Higgs boson production, event 
separation and 
background conditions
are considered. It is shown that the Higgs boson production
process $\gamma e\to
\nu W H$ seems to be very promising for the investigation of
gauge cancellations
between different diagrams and search for anomalous phenomena (for instance,
anomalous Higgs boson interaction vertices). 
  
   In the paper\cite{r5} the calculations of 
   total cross sections 
    for the $W$-
and $Z$-bosons production in $\gamma e$ and $\gamma\gamma$ collisions 
are presented in the 3rd order in
electroweak coupling constant at the tree level. They are compared with the 
estimations obtained
by simple approximation methods to see their accuracy for this class of 
processes.
The preliminary physical analysis of obtained results is given.

    All tree level diagrams for the reaction $\gamma\gamma\to t
\bar{t} H$ have been calculated by means of the
CopmHEP package  \cite{rg}. 
It was shown that the reaction is very sensitive to probing the
Higgs-fermion  coupling in the TeV energy range.  
 
    In the paper\cite{r6} the complete tree level calculations
for three particle final 
state production at  future $e^+e^-, \gamma e$ and $\gamma\gamma$
colliders are presented (see Fig. 5). 
The results obtained with the help of the CompHEP package for total cross 
sections and
other characteristics of processes in the energy range 0.1-2 TeV
are summarized and their comparison with
the results of other approaches is discussed. In particlular 
the processes of $W-,\;Z$- and $H$-bosons production are
considered . These reactions are especially interesting 
in connection with
probing new couplings, searching for new particle signals and estimating
the most important backgrounds in
various experiments.
   
   The possibility of the single and pair excited neutrino production in high
energy $e^+e^-,\; \gamma e$ and $\gamma\gamma$ collisions at linear colliders
 is studied in Ref.\cite{r7}.
The integrated cross sections of these subprocesses are calculated
in a symbolical form. A special attention is paid
to a search for excited neutrino in the $\gamma e\to W W e$
process (see Fig. 6). 
The lower limits for the compositeness parameter to be available
in the experiments at Next Linear Colliders  are estimated.

    The possibility to detect the Higgs boson signal in the process 
$e^+e^-\to Z \bar{b} b$ at LEP200 energies is considered in Ref.\cite{r8}.
 The calculations are performed in the
tree approximation for a complete set of diagrams. Tree level corrections
to the Higgs signal are computed. If the highest possible 
LEPII energy is
$\sqrt{s}$ = 190 GeV the Higgs signal will be very clean for the masses of 
Higgs up to $\sqrt{s}-M_Z$ about 95 GeV.

   In \cite{r9} a possibility of Higgs boson signal observation
   at LEPII and Next Linear Colliders in the reactions
   $e^+e^- \to \mu^+\mu^- b \bar b$, $e^+e^-\to\nu
   \bar{\nu} b \bar{b}$, $e^+e^-\to e^+e^-b \bar b$ is investigated. 
   Complete tree level calculations for these $2 \to 4$ processes
   are performed and compared with the various effective $2 \to 2$
   body approximations. The accuracy of effective approximations near
   the thresholds and at different energies are calculated. In some
   situations it is necessary to introduce nontrivial kinematical
   cuts in order to separate the signal from a background.

A complete tree-level calculation of the reaction 
$e^+e^-\to e \nu \bar{t} b$
in the electroweak standard theory in the LEPII energy range is presented
in Ref.\cite{r10}. For top quark masses
in the range of 130 to 190 GeV the cross sections are found to be of the order
of $10^{-5}-10^{-6}$ pb (see Fig. 7). 
 Therefore, the number of single top quark events is expected to be negligible
even with an integrated luminosity of L=500 pb$^{-1}$. It is further 
demonstrated that the Weizsaecker-Williams
approximation fits the accurate cross section calculations 
reasonably well (see Fig. 8). 

One needs to stress that all results for the reactions with 4 fermions
production in the final state ( Ref.\cite{r9,r10}) have been obtained
taking accout of non-zero masses of all fermions. Such the  complicated 
calculations have been carried out for the first time. Of
course, in some cases 
one can neglect a mass effect using, for instance, cuts on variables
like pair invariant mass or $p_t$. However, sometimes it is really
necessary to keep nonvanishing masses. It is obvious for 
production of such a heavy particle as the top quark. But it happens that 
in order to calculate a total
rate of some reactions one needs to keep even the electron mass 
because the cross section contains a large $\log (s/m_e^2)$. As an example
of such a situation one can demonstrate a calculation of the total rate
of the reaction $e^+e^-\to e^+e^-b \bar b$. You can see in Fig.~9 that
contributions from multiperipheral and single cascade diagrams including the
above mentioned logarithmic terms are significally greeter than from other
subsets of diagrams. In the case when the final
electron and positron go to very forward-backward directions
the reaction $e^+e^-\to e^+e^-b \bar b$  turns out to be a very 
important background
for the Higgs production with the signature $\nu\bar{\nu} b
\bar{b}$. 

Recently based on CompHEP the detail simulation of the
production of light and intermediate Higgs bosons in association with
$W$- and $Z$-bosons have been presented \cite{Higgs-Fnal} at 
TEVATRON collider energies. 
An optimal set of kinematical cuts
to separate the Higgs signal from backgrounds have been found
(see Fig. 10). 
It was shown that one can use such
final states to detect the Higgs signal, however to reduce backgrounds
strongly enough it would be necessary to increase the collider luminosity
up to 500-1000 pb$^{-1}$ and to have an effective $b$-tagging system.

In Ref.\cite{r11} the possibilities of search for vector
leptoquarks at high energy $e p$ and $\gamma p$ 
colliders are investigated. The exact analytical expressions are derived
for cross sections with the help of CompHEP taking account of
possible anomalous couplings of vector leptoquarks with  gauge bosons.
The vector leptoquark search potential of HERA and future $e p$ colliders
is discussed in detail.

    In Ref.\cite{r12} leptoquark resonance production in
electron-proton collisions associated with the emission of a
hard photon is considered.  Estimates of the
sensitivity of the radiative amplitude zero (RAZ) effect (see
Figs. 11,12) 
 as a tool to identify the leptoquarks quantum numbers are carried
out with the help of CompHEP. Also the possibility of measuring the
anomalous couplings of vector leptoquarks with the photon by means of the
RAZ effect is discussed (see Fig. 13). 

\newpage
\begin{center}
{\bf 5. Concluding remarks}
\end{center}

 The main steps of the CompHEP development are published
   in Ref.\cite{r1}.
    Adaptations for various platforms were  done during  visits of the CompHEP
group members to KEK (Japan), Seoul National University (Korea),
University of London (UK),
    University "La Sapienza" (Italy), University of Sao Paulo (Brazil),
     DESY, MPI for physics, University of Karlsruhe (Germany).

The project was supported by Russian 
State Program on High Energy Physics, RFFI (93-02-14428), ISF
(M9B000, M9B300),
   INTAS (93-1180), Japan Society for the Promotion of Science, Japanese
companies KASUMI Co, Ltd and SECOM Co, Ltd. and the Royal
Society of London (UK).

The authors express their deep gratitude to 
all their colleagues and collaborators with whom the main results
have been obtained and discussed.

\newpage

\begin{center}
{\bf Figure captions}
\end{center}
\begin{itemize}

\item[Fig. 1] The  general structure of the CompHEP  package.

\item[Fig. 2] The  menu system of the CompHEP  symbolic part.

\item[Fig. 3] Diagrams for the process $ e^- u\to e^- u Z$ (the LATEX output
of CompHEP).

\item[Fig. 4]  The menu system of the  CompHEP numerical part.

\item[Fig. 5]  Complete tree level calculations
for the cross sections of three particle final 
state production at  future $\gamma e$ and $\gamma\gamma$
colliders in comparison with two particle production processes.

\item[Fig. 6] Invariant mass of outgoing $e^{-}$ and $W^{+}$
distribution for $\gamma e^{-}\to W^{-}W^{+}e^{-}$
process.  Calculations without folding with the photon spectra:  
the dot-dashed line --- invariant mass distribution according to the
Standard Model; the dotted histogram --- distribution when the
excited neutrino is
produced.  The dashed line and solid line histograms represent
corresponding distributions which are folded with
photon spectra.

\item[Fig. 7] Total cross section from the four photon exchange diagrams
          as a function of the electron angular cut for $m_{top}=$ 140 GeV
          and  $\sqrt{s}=$190 GeV calculated.
          The dotted lines correspond to each of the diagrams 
           squared, the solid line is the result for their coherent
          sum.

\item[Fig. 8] Total cross sections for the process       
        $e^+ e^- \rightarrow e^- \bar \nu_e t \bar b\; (e^+ \nu_e \bar t b)$
        as a function of top quark mass for  $\sqrt{s}=$170, 
       190 and 210 GeV. 
            The solid line -- complete tree level calculations,
         the  dashed line -- the subset of photon exchange diagrams 
 only, the dotted line -- the Weizsacker-Williams approximation.

\item[Fig. 9] Effective approximations (dashed lines) and exact 
          calculations (solid lines) corresponding to various subsets of
          diagrams for the process $e^+ e^- \rightarrow e^+ e^- b \bar b$.
          Each subset includes subgraphs of a gauge invariant process.

\item[Fig. 10] The effective $b\bar{b}$ invariant
mass distribution at
$M_{H}=80$~GeV, $\sqrt{s}=2$~TeV  for the cases 
 a) without any cuts  and b) with cuts  applied.
All event numbers correspond to  $L=1000$~pb$^{-1}$.

\item[Fig. 11] Characteristic angular CMS distributions of scalar
leptoquarks in the processes
1) $e^- + d\to\gamma+S^{-1}_3$; 2) $e^+ + u\to\gamma+R_2$;
3) $e^- + u\to\gamma+S_1(S^0_3)$; 4) $e^+ + d\to\gamma+\tilde R_2$.
 The processes 1 and 2 show explicitly the RAZ effect. 
Here $M_{LQ}=300$ GeV,
$\sqrt{s}=304$ GeV and fermion-leptoquark coupling $\lambda=0.3$. 

\item[Fig. 12] Histograms of angular distribution  for
the scalar leptoquark with $M_{LQ}=200$ GeV in the cases of HERA and
future LEPII+LHC colliders. 

\item[Fig. 13] Histograms demonstrating a difference of the RAZ effects for 
the vector leptoquark $V^{-{1\over 2}}_2$ with $M_{LQ}=300$ GeV
in the cases of Yang-Mills photon-leptoquark coupling (left-hand
picture) and minimal coupling (right-hand one). 
\end{itemize}

\vskip 2cm

\newpage

\framebox{
\begin {tabular}{|c|}\hline
                {\bf CompHEP}\\
                {\bf symbolic module}\\ \hline
                \\
                {\sf Lagrangian (SM and beyond)}\\ 
                $\downarrow$\\ 
                {\sf squared matrix element}
                \\
                \\ \hline
\end{tabular}}
\hskip -0.5cm \raise -1.3cm 
\vbox{$\Longrightarrow$\framebox{\it Feynman diagrams}
      \vskip 6 mm
      $\Longrightarrow$\framebox{\it Symbolic answer}
      \vskip 6 mm
      $\Longrightarrow$\framebox{
              \begin{tabular}{c}
                       \it Numerical calculator \\
                       (\large \it cross section, distributions)\\
                       \large \it for 1 $\rightarrow$2, 2 $\rightarrow$2
              \end{tabular}
                                }
     }

\vskip 4mm
\hskip 3.5cm $\Downarrow$
\vskip 4mm

\hskip 1cm \framebox{\begin {tabular}{c}
                         \it FORTRAN code\\
                         \it for squared matrix element
                     \end{tabular}
                    }

\vskip 4mm
\hskip 3.5cm $\Downarrow$
\vskip 4mm

\hskip 1.5cm 
\framebox{
\begin {tabular}{|c|}\hline
          {\bf CompHEP}\\
          {\bf numerical module}\\ \hline
          \\
          {\sf Kinematics}\\
          {\sf Cuts}\\
          {\sf Regularizations}
          \\
          \\ \hline
\end{tabular}}
\hskip -0.5cm  \raise -0.5cm
\vbox{$\Longleftarrow$ \framebox{\sf Structure functions}
   \vskip 1cm
   $\Longleftarrow$\framebox{\begin{tabular}{c}
                                 {\bf Monte-Carlo}\\ \hline
                                 BASES (integration)\\
                                 SPRING (event generation)\\
                             \end{tabular}
                            }
     }

\vskip 4mm
\hskip 1.5cm $\Downarrow$ \hskip 3.5cm $\Downarrow$ 
\vskip 4mm

\hspace*{0.6cm} \framebox{\it Cross section} \hskip 0.5cm 
\framebox{\begin{tabular}{c}
                    \it Event flow \\
                    \it Histograms
          \end{tabular}
         } 

\vskip 1cm
\centerline{Figure 1} 

\newpage

\small
\begin{picture}(400,500)(0,0)

\put(125,465){
\vbox{
\begin {tabular}{|l|}
 \multicolumn{1}{c}{\rm menu 1}\\ \hline
                 QED  \\
                 Fermi model\\ 
                 St. model (unit. gauge)\\ 
                 St. model (Feyn. gauge) \\
                 NEW MODEL \\
                \hline
\end{tabular}

    }
}

\put(200,425){\vector(0,-1){25}}

\put(140,375){
\vbox{
\begin {tabular}{|l|}
\multicolumn{1}{c}{\rm menu2}\\ \hline                  
                 Enter process   \\
                 Edit model  \\ 
                 Delete changes \\ 
                \hline
\end{tabular}

}

}

\put(150,372){\vector(-1,0){80}}
\put(243,382){\vector(1,0){65}} 

\put(0,345){
\begin {tabular}{|l|}
 \multicolumn{1}{c}{\rm menu3}\\ \hline
                 Parameters \\
                 Constraints\\ 
                 Particles\\ 
                 Lagrangian \\
                \hline
\end{tabular}
}

\put(300,367){ 
\begin {tabular}{|l|}
 \multicolumn{1}{c}{\rm menu 4}\\ \hline
                 Squaring  \\
                 View diagrams\\ 
                \hline
\end{tabular}
}

\put(297,370){\line(-1,0){10}}
\put(287,370){\vector(0,-1){50}}
\put(235,265){
\begin {tabular}{|l|}
 \multicolumn{1}{c}{\rm menu 5}\\ \hline
                 View squared diagrams  \\
                 Symbolic calculation\\ 
                 Write results\\ 
                 REDUCE program \\
                 Numerical calculator\\
                 Enter new process\\
                 Interface \\
                \hline
\end{tabular}
}

\put(235,278){\vector(-1,0){150}}
\put(235,250){\line(-1,0){15}}
\put(220,250){\vector(0,-1){45}}

\put(0,250){
\begin {tabular}{|l|}
 \multicolumn{1}{c}{\rm menu 6}\\ \hline
                 FORTRAN code  \\
                 REDUCE code\\ 
                 MATHEMATICA code\\ 
                \hline
\end{tabular}
}

\put(150,165){
\begin {tabular}{|l|}
 \multicolumn{1}{c}{\rm menu 7}\\ \hline
                 View/change data \\
                 Set angular range \\ 
                 Set precision \\
                 Angular dependence \\
                 Parameter dependence\\ 
                \hline
\end{tabular}
}

\put(150,136){\vector(-1,0){55}}
\put(275,149){\vector(1,0){60}}

\put(315,129){
\begin {tabular}{|l|}
 \multicolumn{1}{c}{\rm menu 8}\\ \hline
                 Show plot \\
                 Save results in a file\\ 
                 Recalculate \\ 
                \hline
\end{tabular}
}

\put(0,120){
\begin {tabular}{|l|}
 \multicolumn{1}{c}{\rm menu 9}\\ \hline
                 Total cross section \\
                 Asymmetry \\  
                \hline
\end{tabular}
}

\put(50,100){\vector(0,-1){30}}
\put(0,50){
\begin {tabular}{|l|}
 \multicolumn{1}{c}{\rm menu 10}\\ \hline
                 Show plot   \\
                 Save results in a file\\ 
                \hline
\end{tabular} 
}

\put(0,0){
\centerline{Figure 2} 
}

\end{picture}

\newpage

\def\diagA{
\begin{picture}(170,300)(0,0)
\put(85,5){\makebox(0,0)[b]{diag 1}}
\thicklines
\thinlines
\put(55.0,175.7){\vector(1,0){0}}
\put(34.3,175.7){\makebox(0,0)[r]{$e1$}}
\put(37.1,175.7){\line(1,0){35.7}}
\put(90.7,175.7){\vector(1,0){0}}
\put(90.0,184.3){\makebox(0,0){$e1$}}
\put(72.9,175.7){\line(1,0){35.7}}
\put(147.1,211.4){\makebox(0,0)[l]{$Z$}}
\multiput(108.1,175.7)(4.7,4.7){8}{\rule[-0.5pt]{1.0pt}{1.0pt}}
\multiput(108.6,176.2)(4.7,4.7){8}{\rule[-0.5pt]{1.0pt}{1.0pt}}
\multiput(109.1,176.7)(4.7,4.7){8}{\rule[-0.5pt]{1.0pt}{1.0pt}}
\multiput(109.6,177.2)(4.7,4.7){8}{\rule[-0.5pt]{1.0pt}{1.0pt}}
\multiput(110.1,177.8)(4.7,4.7){8}{\rule[-0.5pt]{1.0pt}{1.0pt}}
\put(126.4,157.9){\vector(1,-1){0}}
\put(147.1,140.0){\makebox(0,0)[l]{$e1$}}
\put(108.6,175.7){\line(1,-1){35.7}}
\put(67.1,140.0){\makebox(0,0)[r]{$A$}}
\multiput(72.4,175.7)(0.0,-4.9){15}{\rule[-0.5pt]{1.0pt}{1.0pt}}
\multiput(72.4,175.2)(0.0,-4.9){15}{\rule[-0.5pt]{1.0pt}{1.0pt}}
\multiput(72.4,174.7)(0.0,-4.9){15}{\rule[-0.5pt]{1.0pt}{1.0pt}}
\multiput(72.4,174.2)(0.0,-4.9){15}{\rule[-0.5pt]{1.0pt}{1.0pt}}
\multiput(72.4,173.7)(0.0,-4.9){15}{\rule[-0.5pt]{1.0pt}{1.0pt}}
\put(55.0,104.3){\vector(1,0){0}}
\put(34.3,104.3){\makebox(0,0)[r]{$u$}}
\put(37.1,104.3){\line(1,0){35.7}}
\put(72.9,104.3){\line(1,0){35.7}}
\put(126.4,86.4){\vector(1,-1){0}}
\put(147.1,68.6){\makebox(0,0)[l]{$u$}}
\put(108.6,104.3){\line(1,-1){35.7}}
\end{picture}
}

\def\diagB{
\begin{picture}(170,300)(0,0)
\put(85,5){\makebox(0,0)[b]{diag 2}}
\thicklines
\thinlines
\put(55.0,175.7){\vector(1,0){0}}
\put(34.3,175.7){\makebox(0,0)[r]{$e1$}}
\put(37.1,175.7){\line(1,0){35.7}}
\put(72.9,175.7){\line(1,0){35.7}}
\put(126.4,193.6){\vector(1,1){0}}
\put(147.1,211.4){\makebox(0,0)[l]{$e1$}}
\put(108.6,175.7){\line(1,1){35.7}}
\put(67.1,140.0){\makebox(0,0)[r]{$A$}}
\multiput(72.4,175.7)(0.0,-4.9){15}{\rule[-0.5pt]{1.0pt}{1.0pt}}
\multiput(72.4,175.2)(0.0,-4.9){15}{\rule[-0.5pt]{1.0pt}{1.0pt}}
\multiput(72.4,174.7)(0.0,-4.9){15}{\rule[-0.5pt]{1.0pt}{1.0pt}}
\multiput(72.4,174.2)(0.0,-4.9){15}{\rule[-0.5pt]{1.0pt}{1.0pt}}
\multiput(72.4,173.7)(0.0,-4.9){15}{\rule[-0.5pt]{1.0pt}{1.0pt}}
\put(55.0,104.3){\vector(1,0){0}}
\put(34.3,104.3){\makebox(0,0)[r]{$u$}}
\put(37.1,104.3){\line(1,0){35.7}}
\put(90.7,104.3){\vector(1,0){0}}
\put(90.0,112.9){\makebox(0,0){$u$}}
\put(72.9,104.3){\line(1,0){35.7}}
\put(147.1,140.0){\makebox(0,0)[l]{$Z$}}
\multiput(108.1,104.3)(4.7,4.7){8}{\rule[-0.5pt]{1.0pt}{1.0pt}}
\multiput(108.6,104.8)(4.7,4.7){8}{\rule[-0.5pt]{1.0pt}{1.0pt}}
\multiput(109.1,105.3)(4.7,4.7){8}{\rule[-0.5pt]{1.0pt}{1.0pt}}
\multiput(109.6,105.8)(4.7,4.7){8}{\rule[-0.5pt]{1.0pt}{1.0pt}}
\multiput(110.1,106.3)(4.7,4.7){8}{\rule[-0.5pt]{1.0pt}{1.0pt}}
\put(126.4,86.4){\vector(1,-1){0}}
\put(147.1,68.6){\makebox(0,0)[l]{$u$}}
\put(108.6,104.3){\line(1,-1){35.7}}
\end{picture}
}
\def\diagC{
\begin{picture}(170,300)(0,0)
\put(85,5){\makebox(0,0)[b]{diag 3}}
\thicklines
\thinlines
\put(72.9,211.4){\vector(1,0){0}}
\put(34.3,211.4){\makebox(0,0)[r]{$e1$}}
\put(37.1,211.4){\line(1,0){71.4}}
\put(126.4,211.4){\vector(1,0){0}}
\put(147.1,211.4){\makebox(0,0)[l]{$e1$}}
\put(108.6,211.4){\line(1,0){35.7}}
\put(102.9,175.7){\makebox(0,0)[r]{$A$}}
\multiput(108.1,211.4)(0.0,-4.9){15}{\rule[-0.5pt]{1.0pt}{1.0pt}}
\multiput(108.1,210.9)(0.0,-4.9){15}{\rule[-0.5pt]{1.0pt}{1.0pt}}
\multiput(108.1,210.4)(0.0,-4.9){15}{\rule[-0.5pt]{1.0pt}{1.0pt}}
\multiput(108.1,209.9)(0.0,-4.9){15}{\rule[-0.5pt]{1.0pt}{1.0pt}}
\multiput(108.1,209.4)(0.0,-4.9){15}{\rule[-0.5pt]{1.0pt}{1.0pt}}
\put(126.4,140.0){\vector(1,0){0}}
\put(147.1,140.0){\makebox(0,0)[l]{$u$}}
\put(108.6,140.0){\line(1,0){35.7}}
\put(108.6,104.3){\vector(0,1){0}}
\put(102.9,104.3){\makebox(0,0)[r]{$u$}}
\put(108.6,140.0){\line(0,-1){71.4}}
\put(72.9,68.6){\vector(1,0){0}}
\put(34.3,68.6){\makebox(0,0)[r]{$u$}}
\put(37.1,68.6){\line(1,0){71.4}}
\put(147.1,68.6){\makebox(0,0)[l]{$Z$}}
\multiput(108.1,68.6)(4.7,0.0){8}{\rule[-0.5pt]{1.0pt}{1.0pt}}
\multiput(108.6,68.6)(4.7,0.0){8}{\rule[-0.5pt]{1.0pt}{1.0pt}}
\multiput(109.1,68.6)(4.7,0.0){8}{\rule[-0.5pt]{1.0pt}{1.0pt}}
\multiput(109.6,68.6)(4.7,0.0){8}{\rule[-0.5pt]{1.0pt}{1.0pt}}
\multiput(110.1,68.6)(4.7,0.0){8}{\rule[-0.5pt]{1.0pt}{1.0pt}}
\end{picture}
}
\def\diagD{
\begin{picture}(170,300)(0,0)
\put(85,5){\makebox(0,0)[b]{diag 4}}
\thicklines
\thinlines
\put(55.0,175.7){\vector(1,0){0}}
\put(34.3,175.7){\makebox(0,0)[r]{$e1$}}
\put(37.1,175.7){\line(1,0){35.7}}
\put(72.9,175.7){\line(1,0){35.7}}
\put(126.4,193.6){\vector(1,1){0}}
\put(147.1,211.4){\makebox(0,0)[l]{$e1$}}
\put(108.6,175.7){\line(1,1){35.7}}
\put(67.1,140.0){\makebox(0,0)[r]{$Z$}}
\multiput(72.4,175.7)(0.0,-4.9){15}{\rule[-0.5pt]{1.0pt}{1.0pt}}
\multiput(72.4,175.2)(0.0,-4.9){15}{\rule[-0.5pt]{1.0pt}{1.0pt}}
\multiput(72.4,174.7)(0.0,-4.9){15}{\rule[-0.5pt]{1.0pt}{1.0pt}}
\multiput(72.4,174.2)(0.0,-4.9){15}{\rule[-0.5pt]{1.0pt}{1.0pt}}
\multiput(72.4,173.7)(0.0,-4.9){15}{\rule[-0.5pt]{1.0pt}{1.0pt}}
\put(55.0,104.3){\vector(1,0){0}}
\put(34.3,104.3){\makebox(0,0)[r]{$u$}}
\put(37.1,104.3){\line(1,0){35.7}}
\put(90.7,104.3){\vector(1,0){0}}
\put(90.0,112.9){\makebox(0,0){$u$}}
\put(72.9,104.3){\line(1,0){35.7}}
\put(147.1,140.0){\makebox(0,0)[l]{$Z$}}
\multiput(108.1,104.3)(4.7,4.7){8}{\rule[-0.5pt]{1.0pt}{1.0pt}}
\multiput(108.6,104.8)(4.7,4.7){8}{\rule[-0.5pt]{1.0pt}{1.0pt}}
\multiput(109.1,105.3)(4.7,4.7){8}{\rule[-0.5pt]{1.0pt}{1.0pt}}
\multiput(109.6,105.8)(4.7,4.7){8}{\rule[-0.5pt]{1.0pt}{1.0pt}}
\multiput(110.1,106.3)(4.7,4.7){8}{\rule[-0.5pt]{1.0pt}{1.0pt}}
\put(126.4,86.4){\vector(1,-1){0}}
\put(147.1,68.6){\makebox(0,0)[l]{$u$}}
\put(108.6,104.3){\line(1,-1){35.7}}
\end{picture}
}
\def\diagE{
\begin{picture}(170,300)(0,0)
\put(85,5){\makebox(0,0)[b]{diag 5}}
\thicklines
\thinlines
\put(72.9,211.4){\vector(1,0){0}}
\put(34.3,211.4){\makebox(0,0)[r]{$e1$}}
\put(37.1,211.4){\line(1,0){71.4}}
\put(126.4,211.4){\vector(1,0){0}}
\put(147.1,211.4){\makebox(0,0)[l]{$e1$}}
\put(108.6,211.4){\line(1,0){35.7}}
\put(102.9,175.7){\makebox(0,0)[r]{$Z$}}
\multiput(108.1,211.4)(0.0,-4.9){15}{\rule[-0.5pt]{1.0pt}{1.0pt}}
\multiput(108.1,210.9)(0.0,-4.9){15}{\rule[-0.5pt]{1.0pt}{1.0pt}}
\multiput(108.1,210.4)(0.0,-4.9){15}{\rule[-0.5pt]{1.0pt}{1.0pt}}
\multiput(108.1,209.9)(0.0,-4.9){15}{\rule[-0.5pt]{1.0pt}{1.0pt}}
\multiput(108.1,209.4)(0.0,-4.9){15}{\rule[-0.5pt]{1.0pt}{1.0pt}}
\put(126.4,140.0){\vector(1,0){0}}
\put(147.1,140.0){\makebox(0,0)[l]{$u$}}
\put(108.6,140.0){\line(1,0){35.7}}
\put(108.6,104.3){\vector(0,1){0}}
\put(102.9,104.3){\makebox(0,0)[r]{$u$}}
\put(108.6,140.0){\line(0,-1){71.4}}
\put(72.9,68.6){\vector(1,0){0}}
\put(34.3,68.6){\makebox(0,0)[r]{$u$}}
\put(37.1,68.6){\line(1,0){71.4}}
\put(147.1,68.6){\makebox(0,0)[l]{$Z$}}
\multiput(108.1,68.6)(4.7,0.0){8}{\rule[-0.5pt]{1.0pt}{1.0pt}}
\multiput(108.6,68.6)(4.7,0.0){8}{\rule[-0.5pt]{1.0pt}{1.0pt}}
\multiput(109.1,68.6)(4.7,0.0){8}{\rule[-0.5pt]{1.0pt}{1.0pt}}
\multiput(109.6,68.6)(4.7,0.0){8}{\rule[-0.5pt]{1.0pt}{1.0pt}}
\multiput(110.1,68.6)(4.7,0.0){8}{\rule[-0.5pt]{1.0pt}{1.0pt}}
\end{picture}
}
\def\diagF{
\begin{picture}(170,300)(0,0)
\put(85,5){\makebox(0,0)[b]{diag 6}}
\thicklines
\thinlines
\put(55.0,175.7){\vector(1,0){0}}
\put(34.3,175.7){\makebox(0,0)[r]{$e1$}}
\put(37.1,175.7){\line(1,0){35.7}}
\put(90.7,175.7){\vector(1,0){0}}
\put(90.0,184.3){\makebox(0,0){$e1$}}
\put(72.9,175.7){\line(1,0){35.7}}
\put(147.1,211.4){\makebox(0,0)[l]{$Z$}}
\multiput(108.1,175.7)(4.7,4.7){8}{\rule[-0.5pt]{1.0pt}{1.0pt}}
\multiput(108.6,176.2)(4.7,4.7){8}{\rule[-0.5pt]{1.0pt}{1.0pt}}
\multiput(109.1,176.7)(4.7,4.7){8}{\rule[-0.5pt]{1.0pt}{1.0pt}}
\multiput(109.6,177.2)(4.7,4.7){8}{\rule[-0.5pt]{1.0pt}{1.0pt}}
\multiput(110.1,177.8)(4.7,4.7){8}{\rule[-0.5pt]{1.0pt}{1.0pt}}
\put(126.4,157.9){\vector(1,-1){0}}
\put(147.1,140.0){\makebox(0,0)[l]{$e1$}}
\put(108.6,175.7){\line(1,-1){35.7}}
\put(67.1,140.0){\makebox(0,0)[r]{$Z$}}
\multiput(72.4,175.7)(0.0,-4.9){15}{\rule[-0.5pt]{1.0pt}{1.0pt}}
\multiput(72.4,175.2)(0.0,-4.9){15}{\rule[-0.5pt]{1.0pt}{1.0pt}}
\multiput(72.4,174.7)(0.0,-4.9){15}{\rule[-0.5pt]{1.0pt}{1.0pt}}
\multiput(72.4,174.2)(0.0,-4.9){15}{\rule[-0.5pt]{1.0pt}{1.0pt}}
\multiput(72.4,173.7)(0.0,-4.9){15}{\rule[-0.5pt]{1.0pt}{1.0pt}}
\put(55.0,104.3){\vector(1,0){0}}
\put(34.3,104.3){\makebox(0,0)[r]{$u$}}
\put(37.1,104.3){\line(1,0){35.7}}
\put(72.9,104.3){\line(1,0){35.7}}
\put(126.4,86.4){\vector(1,-1){0}}
\put(147.1,68.6){\makebox(0,0)[l]{$u$}}
\put(108.6,104.3){\line(1,-1){35.7}}
\end{picture}
}
\def\diagG{
\begin{picture}(170,300)(0,0)
\put(85,5){\makebox(0,0)[b]{diag 7}}
\thicklines
\thinlines
\put(72.9,211.4){\vector(1,0){0}}
\put(34.3,211.4){\makebox(0,0)[r]{$e1$}}
\put(37.1,211.4){\line(1,0){71.4}}
\put(147.1,211.4){\makebox(0,0)[l]{$Z$}}
\multiput(108.1,211.4)(4.7,0.0){8}{\rule[-0.5pt]{1.0pt}{1.0pt}}
\multiput(108.6,211.4)(4.7,0.0){8}{\rule[-0.5pt]{1.0pt}{1.0pt}}
\multiput(109.1,211.4)(4.7,0.0){8}{\rule[-0.5pt]{1.0pt}{1.0pt}}
\multiput(109.6,211.4)(4.7,0.0){8}{\rule[-0.5pt]{1.0pt}{1.0pt}}
\multiput(110.1,211.4)(4.7,0.0){8}{\rule[-0.5pt]{1.0pt}{1.0pt}}
\put(108.6,175.7){\vector(0,-1){0}}
\put(102.9,175.7){\makebox(0,0)[r]{$e1$}}
\put(108.6,211.4){\line(0,-1){71.4}}
\put(126.4,140.0){\vector(1,0){0}}
\put(147.1,140.0){\makebox(0,0)[l]{$e1$}}
\put(108.6,140.0){\line(1,0){35.7}}
\put(102.9,104.3){\makebox(0,0)[r]{$A$}}
\multiput(108.1,140.0)(0.0,-4.9){15}{\rule[-0.5pt]{1.0pt}{1.0pt}}
\multiput(108.1,139.5)(0.0,-4.9){15}{\rule[-0.5pt]{1.0pt}{1.0pt}}
\multiput(108.1,139.0)(0.0,-4.9){15}{\rule[-0.5pt]{1.0pt}{1.0pt}}
\multiput(108.1,138.5)(0.0,-4.9){15}{\rule[-0.5pt]{1.0pt}{1.0pt}}
\multiput(108.1,138.0)(0.0,-4.9){15}{\rule[-0.5pt]{1.0pt}{1.0pt}}
\put(72.9,68.6){\vector(1,0){0}}
\put(34.3,68.6){\makebox(0,0)[r]{$u$}}
\put(37.1,68.6){\line(1,0){71.4}}
\put(126.4,68.6){\vector(1,0){0}}
\put(147.1,68.6){\makebox(0,0)[l]{$u$}}
\put(108.6,68.6){\line(1,0){35.7}}
\end{picture}
}
\def\diagH{
\begin{picture}(170,300)(0,0)
\put(85,5){\makebox(0,0)[b]{diag 8}}
\thicklines
\thinlines
\put(72.9,211.4){\vector(1,0){0}}
\put(34.3,211.4){\makebox(0,0)[r]{$e1$}}
\put(37.1,211.4){\line(1,0){71.4}}
\put(147.1,211.4){\makebox(0,0)[l]{$Z$}}
\multiput(108.1,211.4)(4.7,0.0){8}{\rule[-0.5pt]{1.0pt}{1.0pt}}
\multiput(108.6,211.4)(4.7,0.0){8}{\rule[-0.5pt]{1.0pt}{1.0pt}}
\multiput(109.1,211.4)(4.7,0.0){8}{\rule[-0.5pt]{1.0pt}{1.0pt}}
\multiput(109.6,211.4)(4.7,0.0){8}{\rule[-0.5pt]{1.0pt}{1.0pt}}
\multiput(110.1,211.4)(4.7,0.0){8}{\rule[-0.5pt]{1.0pt}{1.0pt}}
\put(108.6,175.7){\vector(0,-1){0}}
\put(102.9,175.7){\makebox(0,0)[r]{$e1$}}
\put(108.6,211.4){\line(0,-1){71.4}}
\put(126.4,140.0){\vector(1,0){0}}
\put(147.1,140.0){\makebox(0,0)[l]{$e1$}}
\put(108.6,140.0){\line(1,0){35.7}}
\put(102.9,104.3){\makebox(0,0)[r]{$Z$}}
\multiput(108.1,140.0)(0.0,-4.9){15}{\rule[-0.5pt]{1.0pt}{1.0pt}}
\multiput(108.1,139.5)(0.0,-4.9){15}{\rule[-0.5pt]{1.0pt}{1.0pt}}
\multiput(108.1,139.0)(0.0,-4.9){15}{\rule[-0.5pt]{1.0pt}{1.0pt}}
\multiput(108.1,138.5)(0.0,-4.9){15}{\rule[-0.5pt]{1.0pt}{1.0pt}}
\multiput(108.1,138.0)(0.0,-4.9){15}{\rule[-0.5pt]{1.0pt}{1.0pt}}
\put(72.9,68.6){\vector(1,0){0}}
\put(34.3,68.6){\makebox(0,0)[r]{$u$}}
\put(37.1,68.6){\line(1,0){71.4}}
\put(126.4,68.6){\vector(1,0){0}}
\put(147.1,68.6){\makebox(0,0)[l]{$u$}}
\put(108.6,68.6){\line(1,0){35.7}}
\end{picture}
}
{
\unitlength=0.5pt
\scriptsize
\diagA\ \diagB\ \diagC\ \diagD\

\diagE\ \diagF\ \diagG\ \diagH
\normalsize
\begin{center}
{Figure 3} 
\end{center}
}

\newpage

\bf

\vbox{
\centerline{
\begin {tabular}{|ll|}
 \multicolumn{2}{c}{\rm Main menu }\\ \hline
1. Calculation      &  2. IN state \\
3. Model parameters &  4. Invariant cuts \\
5. Kinematics       &  6. MC parameters \\
7. Regularization   &   8. Task formation \\
9. View results     & 10. User`s menu \\
 \hline
\end{tabular}
}
  }

\centerline{
\begin{picture}(160,100)
\put(0,20){
\vbox{ 
\begin {tabular}{|l|}
 \multicolumn{1}{c}{\rm In state }\\ \hline
1. StructF(1) = {\it OFF } \\    2. SQRTS = {\it 1000 }   \\
3. StructF(2) = {\it OFF }                 \\
 \hline
\end{tabular}
   }
   }
\end{picture}
\begin{picture}(160,100)  
\put(0,20){
\vbox{
\begin {tabular}{|l|}
\multicolumn{1}{c}{\rm Invariant cuts}\\ \hline 
1. Insert new cut \\ 2. Delete cut \\
3. Change cut        \\
 \hline
\end{tabular}
  }
  }
\end{picture}
}

\vskip 1.0 cm
\vbox{
\centerline{
\begin {tabular}{|ll|}
 \multicolumn{2}{c}{\rm MC parameters }\\ \hline
1. Ncall = {\it 10000} &  2. Acc1= {\it 0.1 }\\
3. Itmx1= {\it 5 }   & 4. Acc2= {\it 0.1 } \\ 
5. Itmx2={\it 0}   & 6. Event gererator {\rm OFF} \\
7. Number of events = {\it 1000 } &  \\
 \hline
\end{tabular}
  }
}

\centerline{
\begin{picture}(180,100)
\put(0,20){
\vbox{
\begin {tabular}{|l|}
 \multicolumn{1}{c}{\rm Regularization }\\ \hline
1. Insert new regularization \\  2. Delete regularization \\
3. Change regularization       \\
 \hline
\end{tabular}
}
}
\end{picture}
\begin{picture}(160,100)
\put(0,20){
\vbox{
\begin {tabular}{|l|}
 \multicolumn{1}{c}{\rm Task formation }\\ \hline
1. Table parameters  \\  2. Set default session \\
3. Add session to batch  \\
 \hline
\end{tabular}
}
}
\end{picture}
}
\vskip 1.0 cm
\vbox{
\centerline{
\begin {tabular}{|ll|}
 \multicolumn{2}{c}{\rm View results}\\ \hline
1.  session \# to view - {\it 3 } &  2. View result file \\
3. View protocol file      &  4. View histogram file \\
 \hline
\end{tabular}
  }
}
\vskip 1.0 cm
{
\centerline{\rm Figure 4}
}


\unitlength=1cm

\begin{figure}[hp]
\begin{picture}(10,8)
\put(-1.5,-0.5){\epsfxsize=17cm \epsfysize=10cm \leavevmode \epsfbox{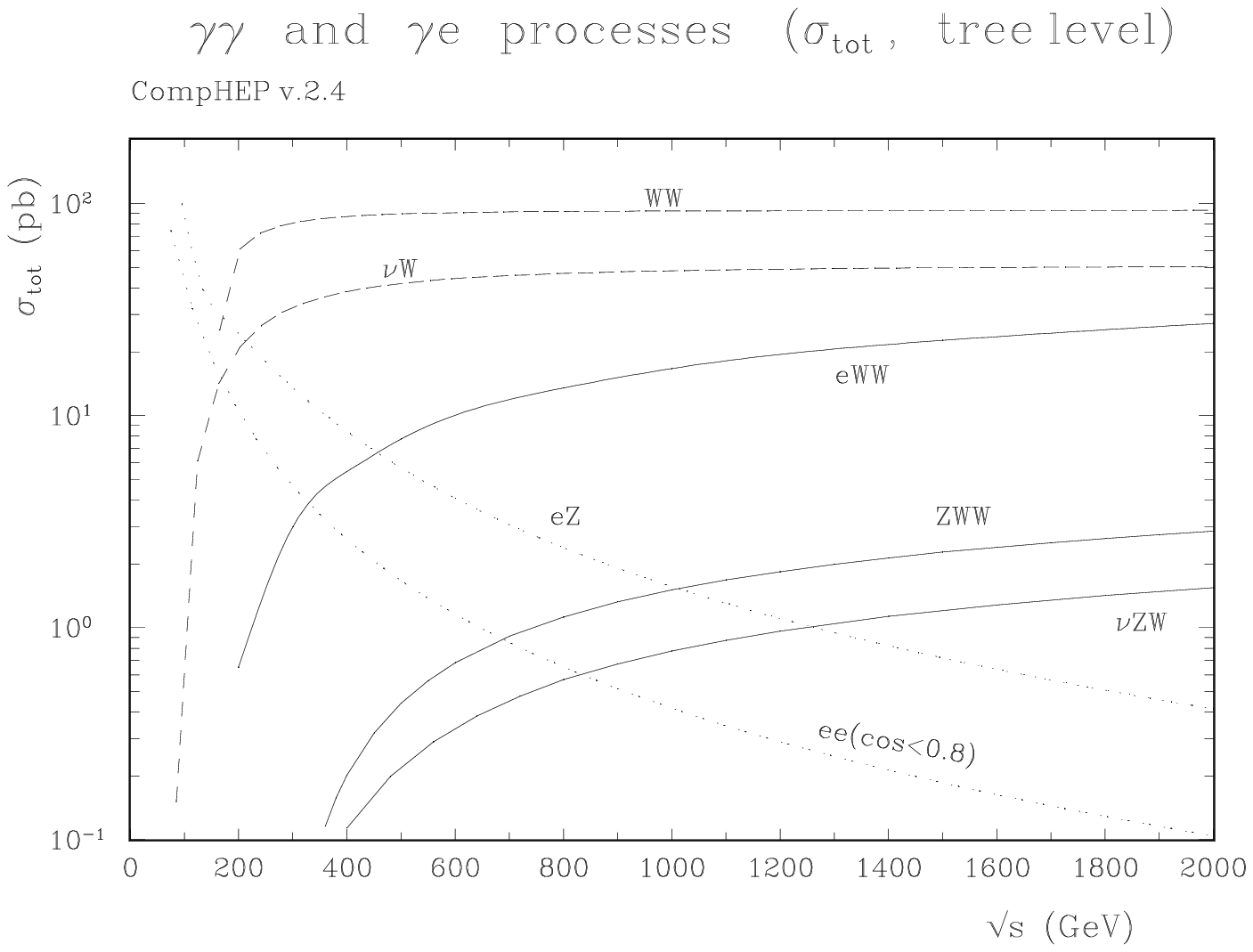} }
\put(7,-1){Figure 5}
\end{picture}
\end{figure}





\vskip 3.5cm

\begin{figure}[hp]
\begin{picture}(10,8)
\put(1,0){\epsfxsize=24cm \epsfysize=10cm \leavevmode \epsfbox{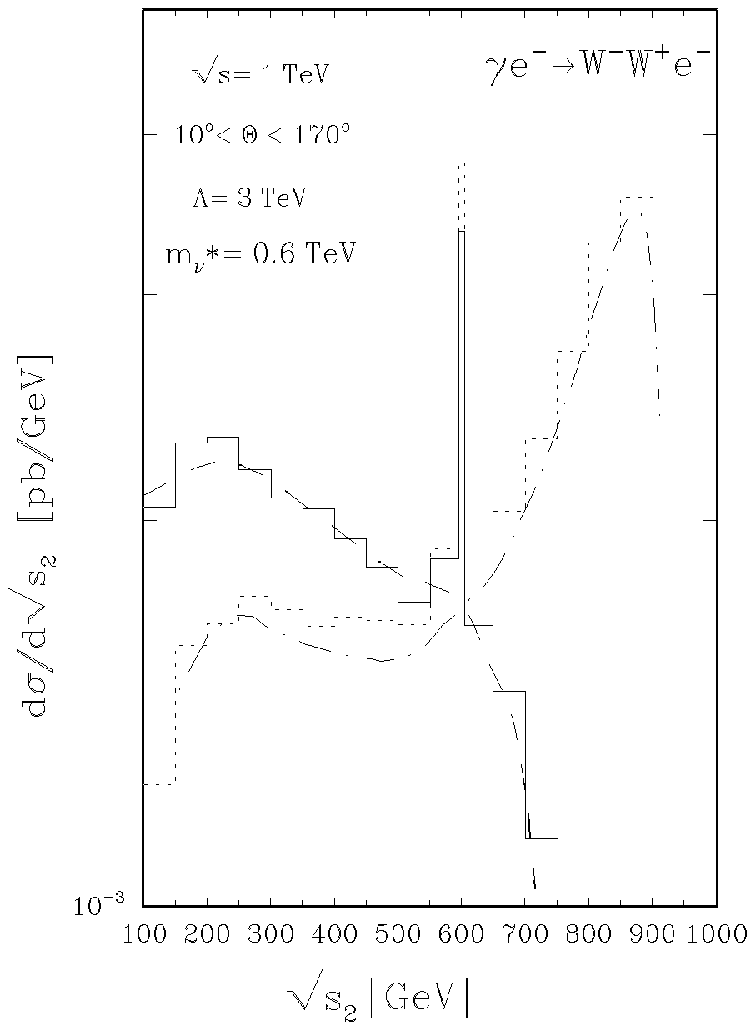} }
\put(7,-1.5){Figure 6}
\end{picture}
\end{figure}


\newpage



\begin{figure}[hp]
\begin{picture}(8,8)
\put(0,-4){\epsfxsize=13cm \epsfysize=11cm \leavevmode \epsfbox{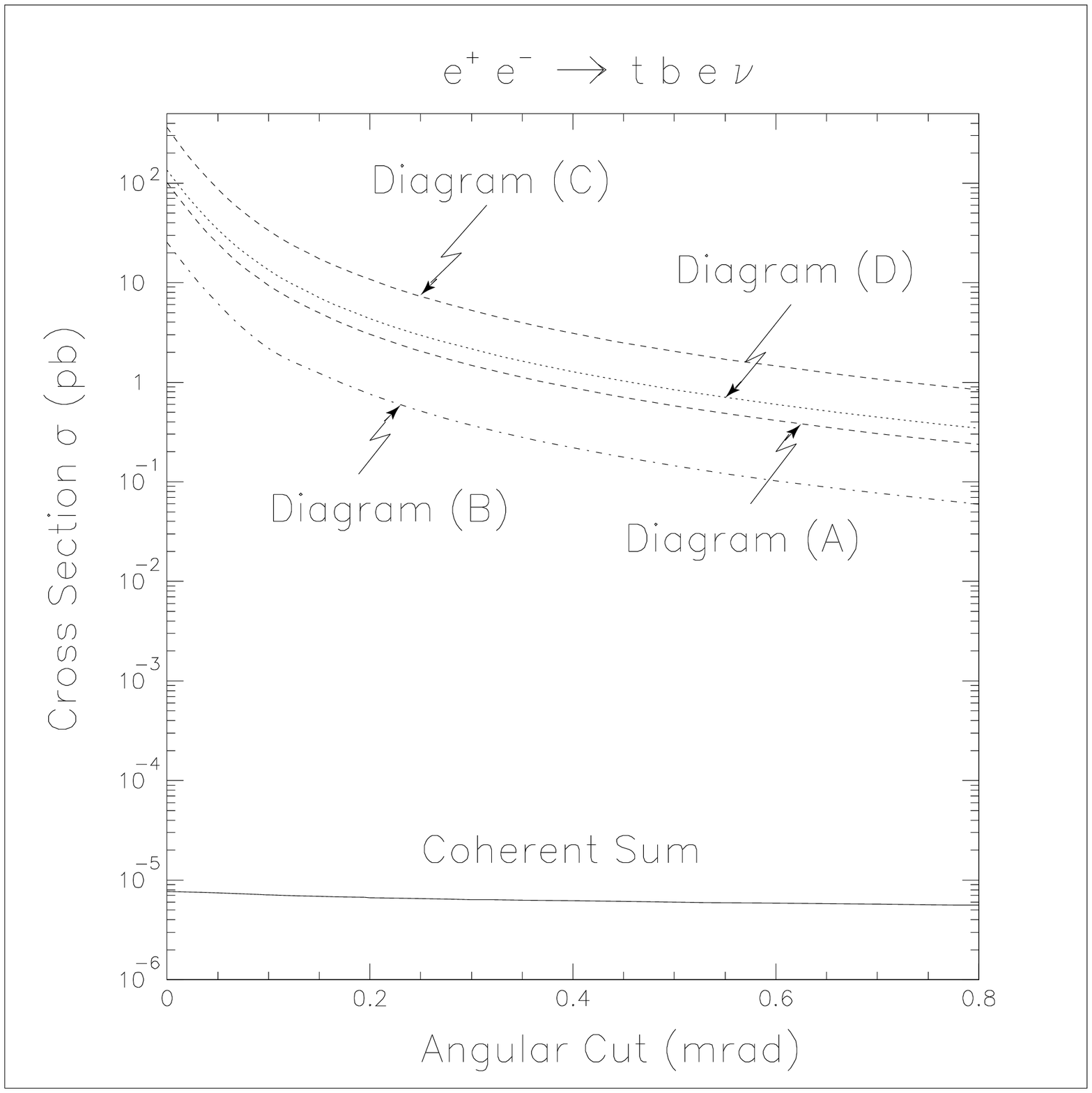} }
\put(7,-2.8){Figure 7}
\end{picture}
\end{figure}



\begin{figure}[hp]
\begin{picture}(8,8)
\put(0,-5.5){\epsfxsize=14cm \epsfysize=11cm \leavevmode \epsfbox{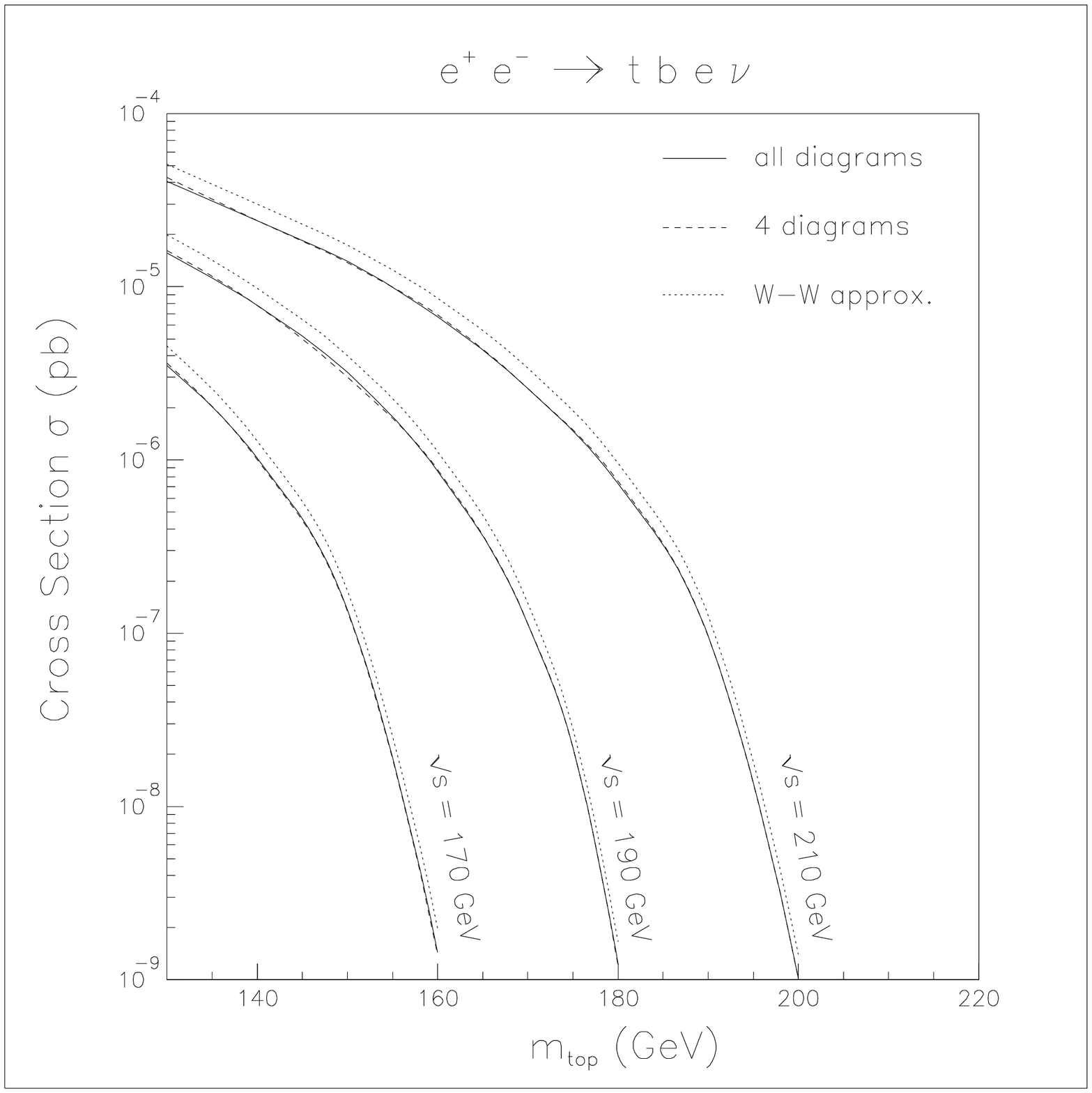} }
\put(7,-4.8){Figure 8}
\end{picture}
\end{figure}
	
\newpage

\unitlength=10mm

\begin{figure}[tbp]
\begin{picture}(10,10)
\put(0,-0.5){\epsfxsize=14cm \epsfysize=10cm \leavevmode \epsfbox{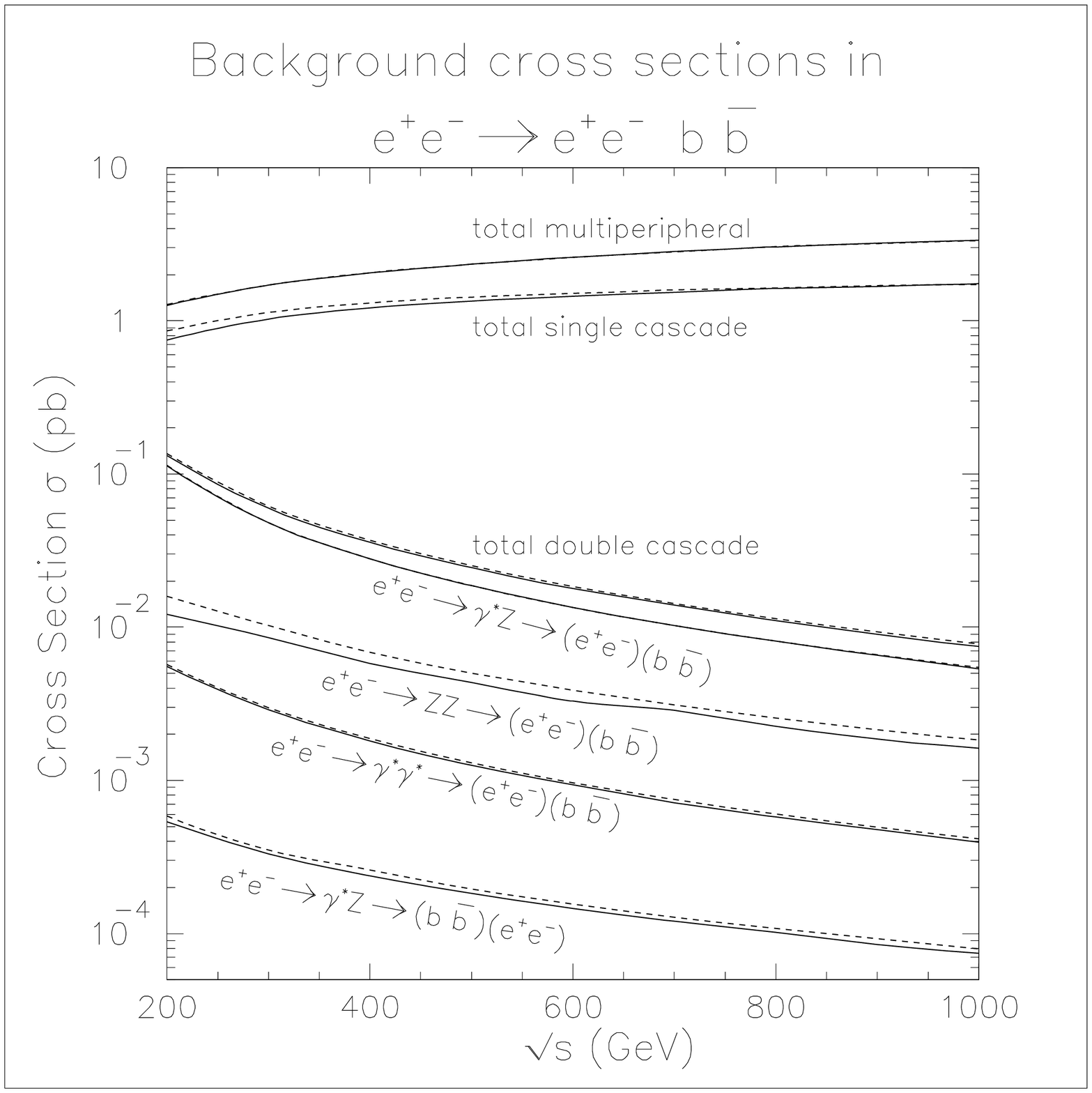} }
\put(7,-1.5){Figure 9}
\end{picture}

\end{figure}




\begin{figure}[tbp]
\begin{picture}(10,10)
\put(0.5,-1.5){\epsfxsize=14cm \epsfysize=11cm \leavevmode \epsfbox{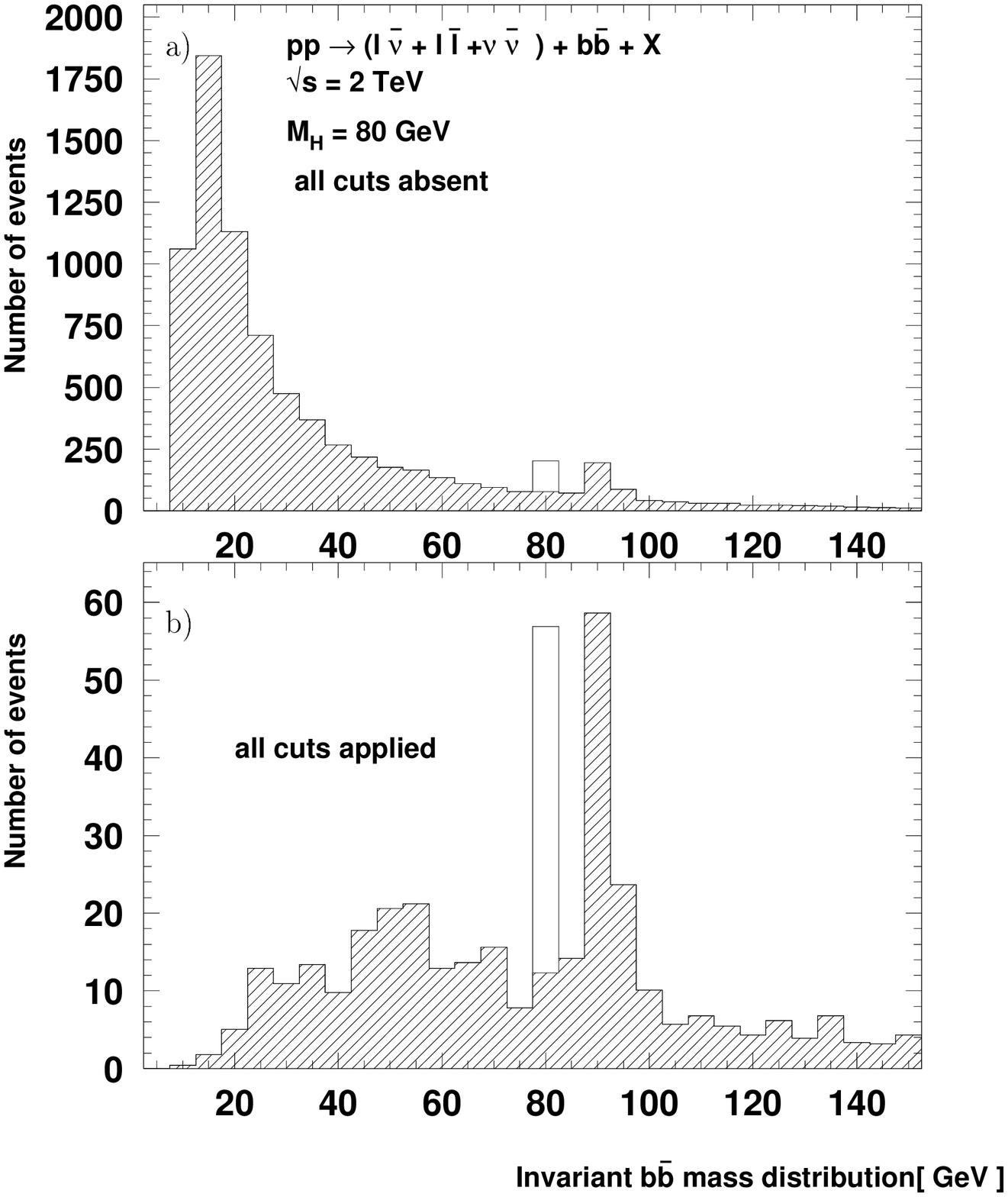} }
\put(7,-0.5){Figure 10}
\end{picture}

\end{figure}

\newpage
\unitlength=1cm

\begin{figure}[h]
\begin{picture}(16,6)
\put(2,0){\epsfxsize=11cm \epsfysize=6cm \leavevmode \epsfbox{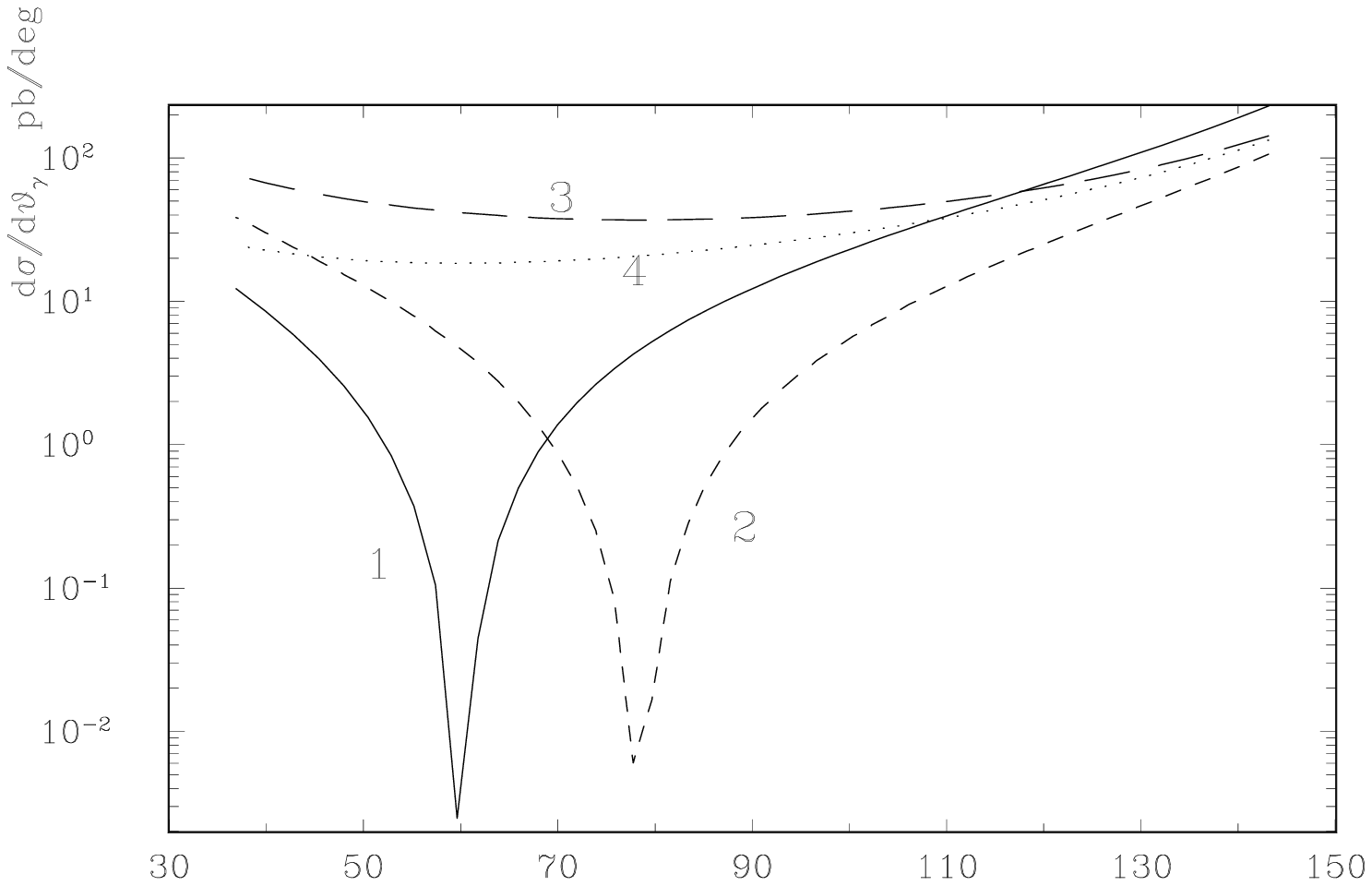} }
\put(7,-0.5){Figure 11}
\end{picture}
\end{figure}

\begin{figure}[h]
\begin{picture}(16,6)
\put(-0.5,1){\epsfxsize=7.5cm \leavevmode \epsfbox{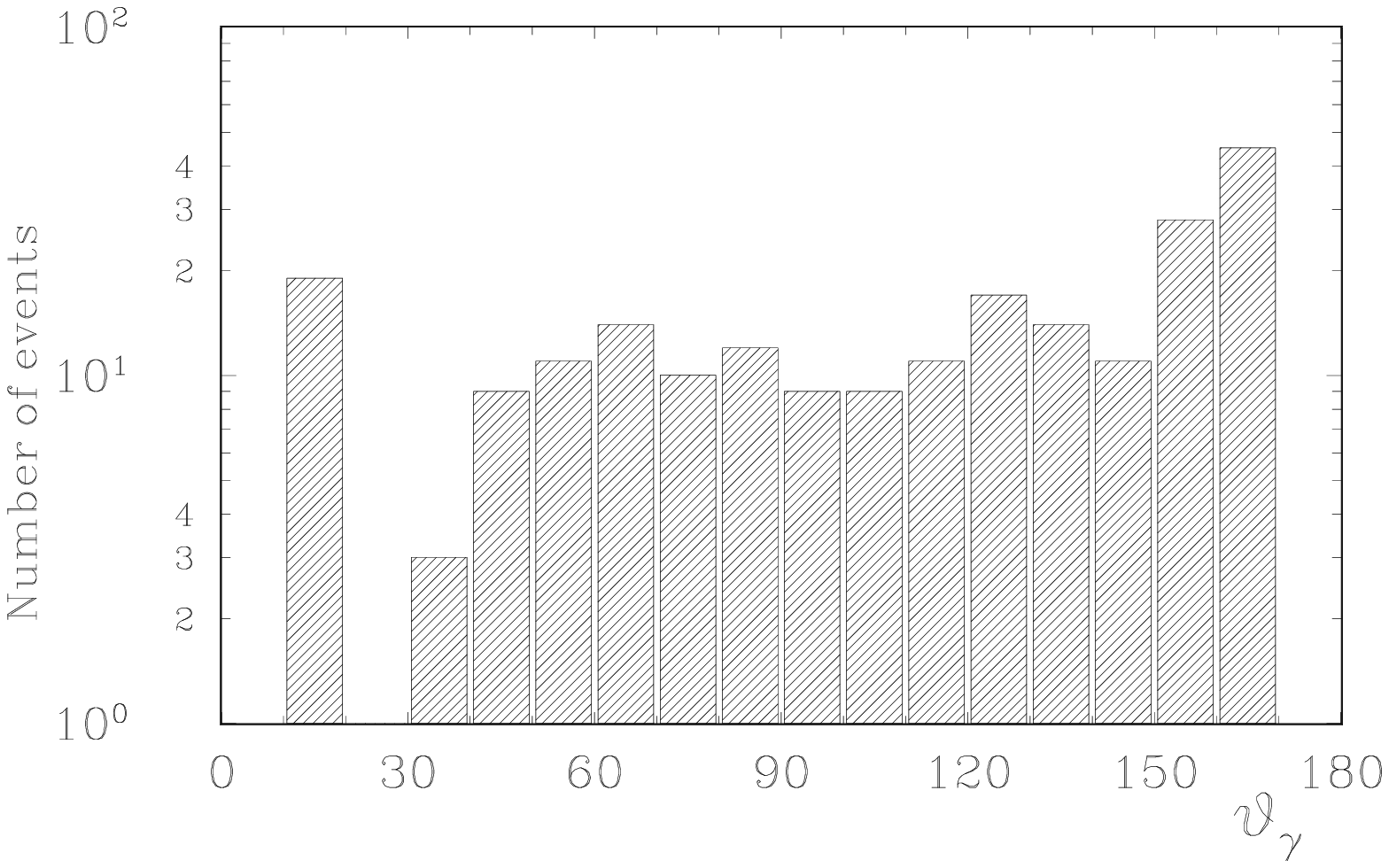} }
\put(1.5,5){\scriptsize $e^++u\rightarrow \gamma+ R_2^{-\frac{1}{2}} 
\quad HERA$} 
\put(9.3,4.8){\scriptsize $e^++u\rightarrow \gamma+ R_2^{-\frac{1}{2}}
 \quad LEP+LHC$}
\put(7.5,1){\epsfxsize=7.5cm \leavevmode \epsfbox{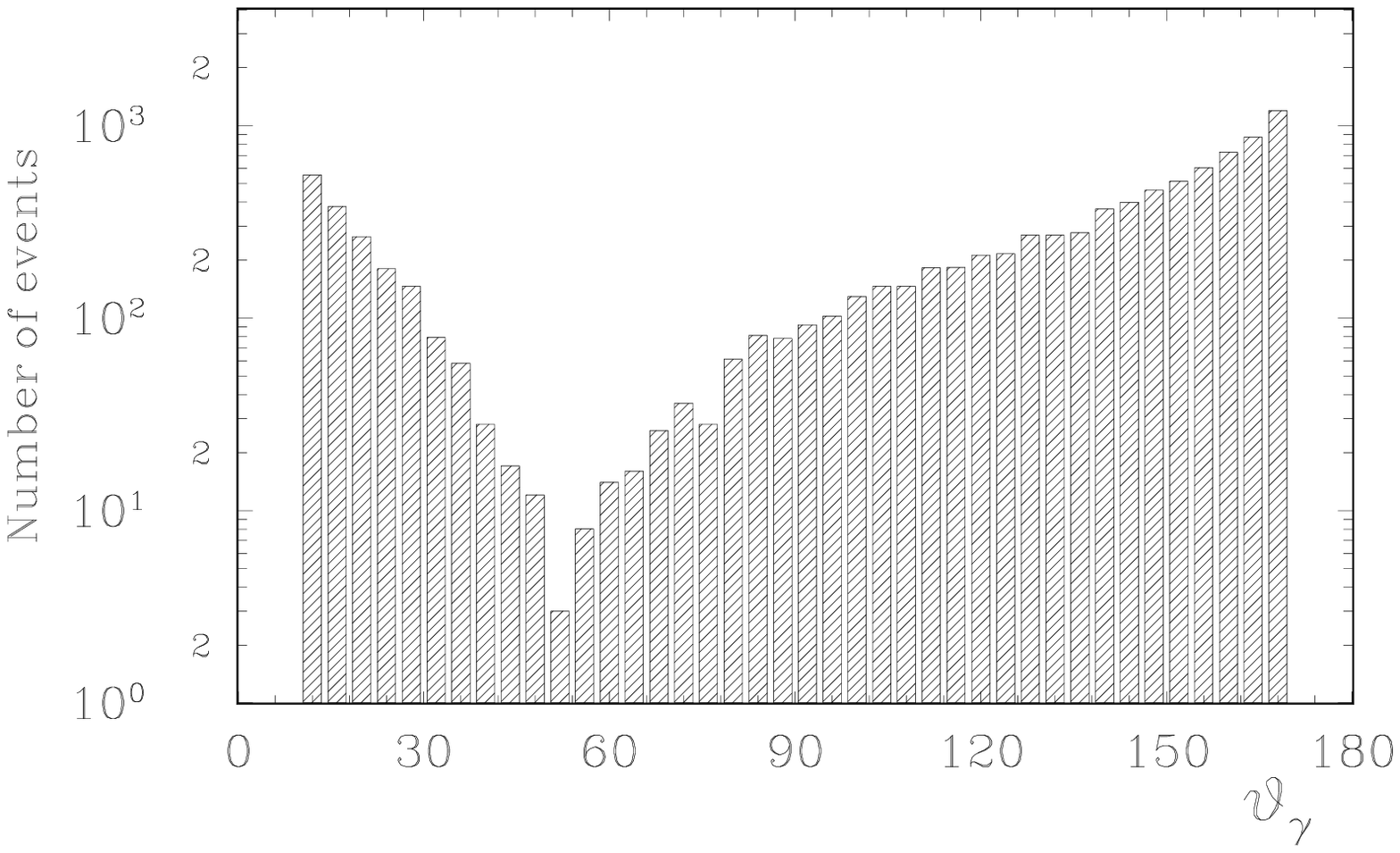} }
\put(7,0.5){Figure 12}
\end{picture}
\end{figure}

\begin{figure}[h]
\begin{picture}(16,6)
\put(-0.5,3){\epsfxsize=7.5cm \leavevmode \epsfbox{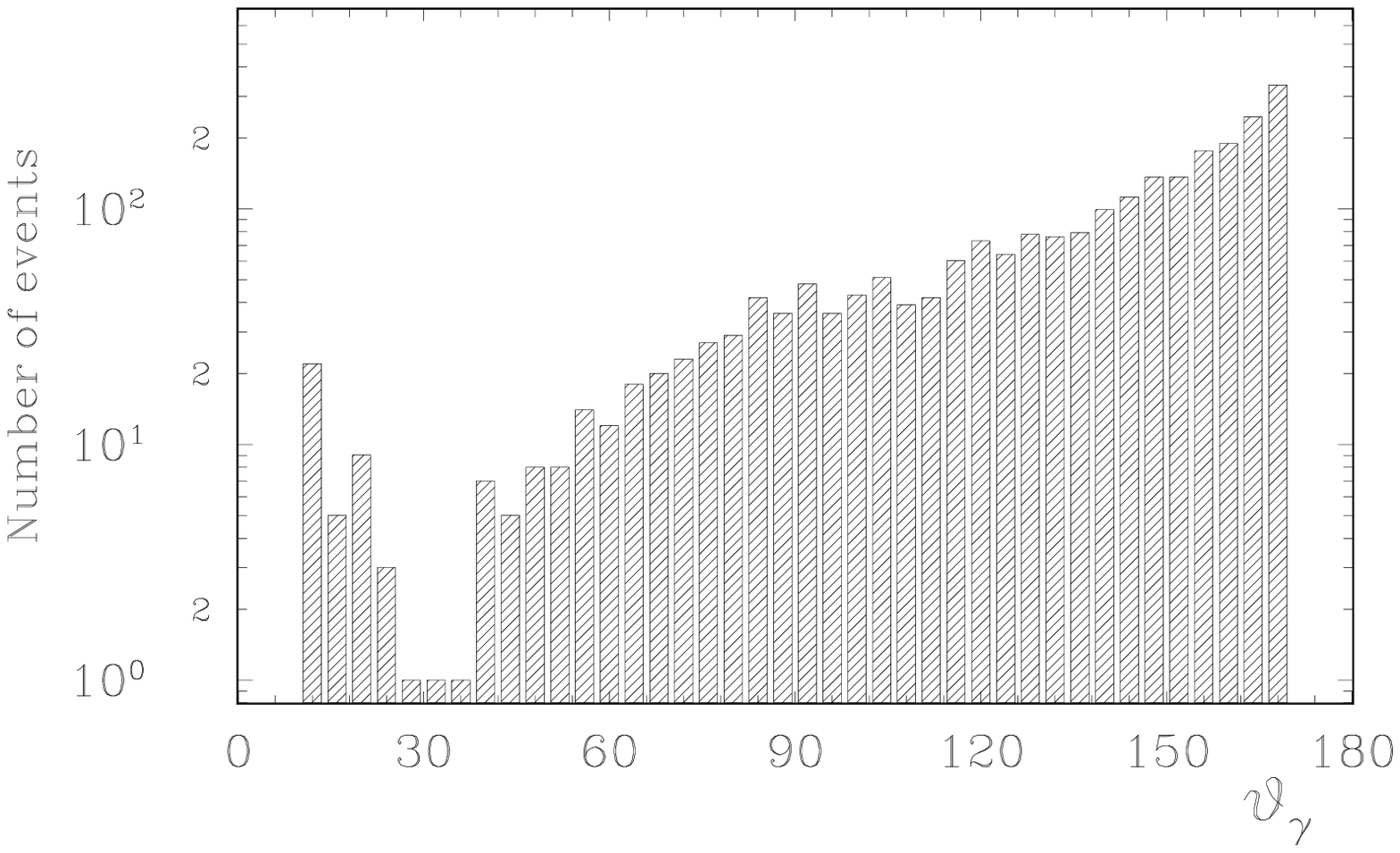}}
\put(7.5,3){\epsfxsize=7.5cm \leavevmode \epsfbox{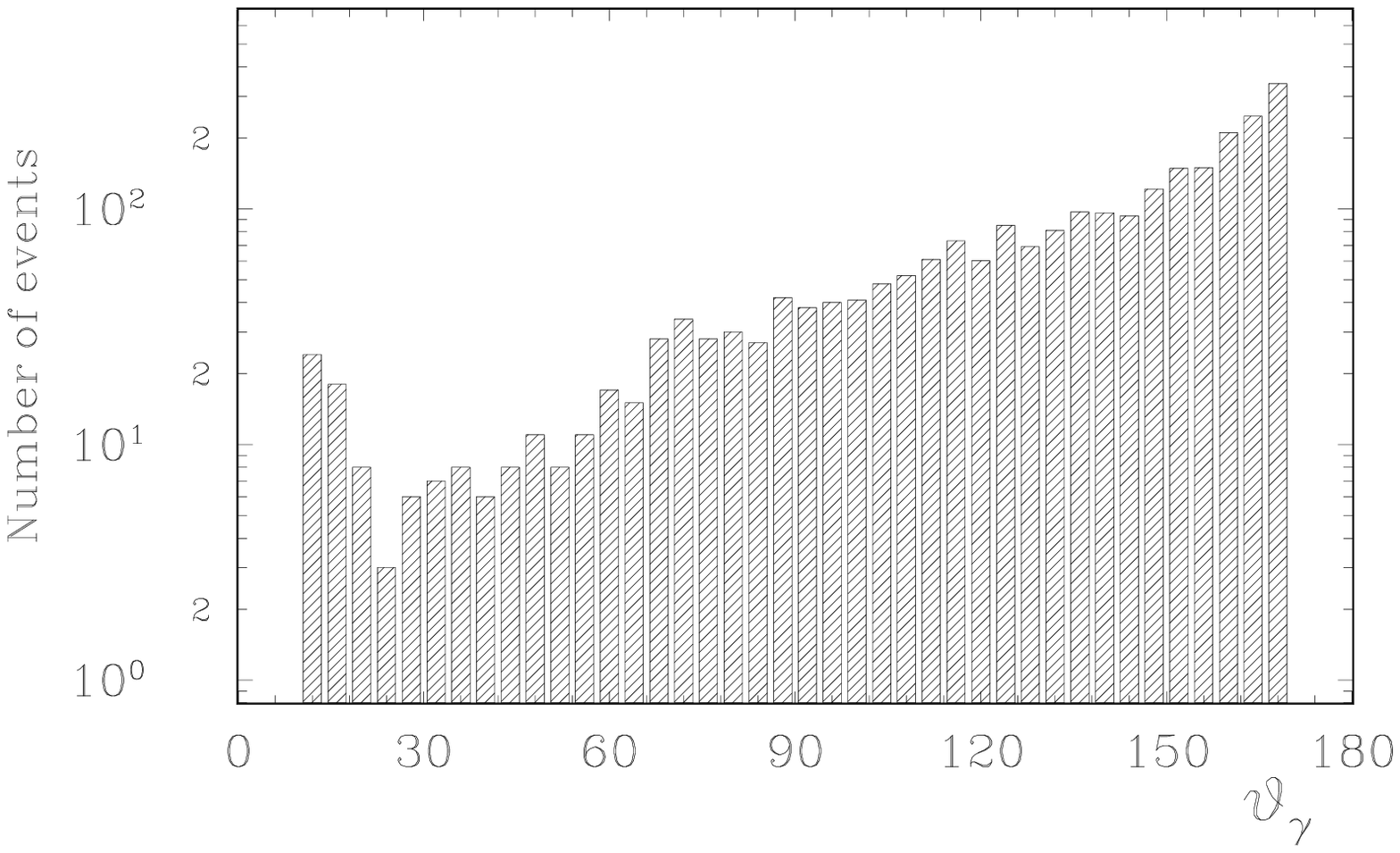}}
\put(7,2.5){Figure 13}
\end{picture}
\end{figure}



\vspace{1cm}

\begin{thebibliography}{99}

\bibitem{r1} E. Boos, M. Dubinin, V. Edneral, V. Ilyin, A. Kryukov, A.Pukhov,
    V. Savrin, S. Shichanin and A. Taranov.  
    in: Proc. Int. Workshop on Software Engineering, Artificial Intelligence
    and Expert Systems for High Energy and Nuclear Physics (AI'90), ed.
    by D. Perret-Gallix and W. Wojcik, Editions du CNRS, 1990, p. 573.   

    L. Gladilin, V. Ilyin, A. Pukhov.   
    in: Proc. CHEP92, ed.by C.Verkerk and W.Wojcik, CERN 92-07, 1992, p. 855.

    E. Boos, M. Dubinin, V. Edneral, V. Ilyin, A. Kryukov, A. Pukhov, 
    S. Shichanin.   
    in: "New Computing Techniques in Physics Research II", 
    ed.by D. Perret-Gallix, World Scientific, Singapore, 1992, p. 665.

    A. Pukhov.  
    in: "New Computing Techniques in Physics Research III", 
    ed. by K.-H.Becks and D.Perret-Gallix, World Scientific, 
    Singapore, 1993, p. 473.

    E. Boos, M. Dubinin, V. Ilyin, A. Pukhov, V. Savrin. Preprint SNUTP
94-116 and INP MSU - 94-36/358, Seoul--Moscow, 1994. 
 

\bibitem{r2} S. Kawabata.  
    Comp. Phys. Comm. 41 (1986) 127.  
    
    S. Kawabata, T. Kaneko.  
    Comp. Phys. Comm. 48 (1988) 353.
    
    T. Ishikawa, T. Kaneko,K. Kato, S. Kawabata, Y. Shimizu and
    H. Tanaka.  
    GRACE manual, KEK report 92-19, 1993.
    
    
\bibitem{r3} E. Boos, M. Dubinin, V. Ilyin, A. Pukhov, S. Shichanin, 
    Y. Shimizu, T. Kaneko, S. Kawabata, Y. Kurihara.  
    Int. J. Mod. Phys. C5 (1994) 615-628.  
     
\bibitem{rCuypers} M. Baillargeon, F. Boudjema, F. Cuypers, E. Gabrielli,
                   B. Mele.
                   Nucl.Phys. B424 (1994) 343
       
\bibitem{r4} E. Boos, M. Dubinin, V. Ilyin, A. Pukhov, G. Jikia, S. 
Sultanov. Phys.Lett. B273 (1991) 173.   
    
  

\bibitem{r5} I. Ginzburg, V. Ilyin, A. Pukhov, V. Serbo, S. Shichanin.   
   Phys. of Atomic Nuclei, 56 (1993) 1481 (Yad.Fiz., 56 (1993) 57)  
    
\bibitem{rg} E. Boos, I. Ginzburg, K. Melnikov, T. Sack, S. Shichanin.
    Z.Phys. C56 (1992) 487.


\bibitem{r6} E. Boos, M. Dubinin, V. Ilyin, A. Pukhov.  
   in: $e^+e^-$ collisions at 500 GeV: The physics potential. Proc. of the
    Workshop
   Munich-Annecy-Hamburg, ed. by P.M.Zerwas, 
   DESY report 93-123C, 1993, p. 561.  
   

\bibitem{r7}A. Belyaev, E. Boos, A. Pukhov.  
   Phys.Lett., B296 (1992) 452.  
     

   A. Belyaev, E. Boos.  
  Phys. of Atomic Nuclei, 56 (1993) 1447 (Yad.Fiz., 56 (1993) 5)  
   

\bibitem{r8} E. Boos, M. Dubinin.   
   Phys. Lett. B308 (1993) 147.

             E.Boos, M.Dubinin.
   Phys. of Atomic Nuclei, 56 (1993) 1455 (Yad.Fiz., 56 (1993) 16)  
   

\bibitem{r9} E. Boos, M. Sachwitz, H. J. Schreiber, S. Shichanin.   
   Z.Phys. C61 (1994) 675. 

             M. Dubinin, V. Edneral, Y. Kurihara, Y. Shimizu.
   Phys.Lett., B329 (1994) 379

             E. Boos, M. Sachwitz, H. J. Schreiber, S. Shichanin.
   Int. J. Mod. Phys A, 10 (1995) 2067 


             E. Boos, M. Sachwitz, H. J. Schreiber, S. Shichanin.
   DESY preprint 94-091 (1994), Z. Phys. C (to be published)

\bibitem{r10} E. Boos,  V. Ilyin, A. Pukhov, S. Shichanin, M. Sachwitz, 
   H. J. Schreiber,
   T. Ishikawa, T. Kaneko, S. Kawabata, Y. Kurihara and Y. Shimizu.  
   Phys.Lett., B326 (1994) 190.  


\bibitem{Higgs-Fnal}
A.Belyaev, E.Boos, L.Dudko
   Mod. Phys. Lett. A, 10 (1995) 25



\bibitem{r11} J. Blumlein, E. Boos, A. Pukhov. 
   Mod. Phys. Lett. A, 9 (1994) 3007
   
    \bibitem{r12} V.A. Ilyin et al., Phys. Lett., B351 (1995)
504; Phys. Lett., B356 (1995)
531;\\ Preprint INP MSU 95-22/386 \& QMW-PH-95-22, 1995, Moscow-London.
\end{thebibliography}
\end{document}